# Quenching of Cross Sections in Nucleon Transfer Reactions

B. P. Kay,[1, 2, ∗] J. P. Schiffer,[1] and S. J. Freeman[3]

[1]*Physics Division, Argonne National Laboratory, Argonne, Illinois 60439, USA*
[2]*Department of Physics, University of York, Heslington, York YO10 5DD, United Kingdom*
[3]*School of Physics and Astronomy, University of Manchester, Manchester M13 9PL, United Kingdom*

(Dated: July 3, 2013)

Cross sections for proton knockout observed in $(e,e'p)$ reactions are apparently quenched by a factor of ∼0.5, an effect attributed to short-range correlations between nucleons. Here we demonstrate that such quenching is not restricted to proton knockout, but a more general phenomenon associated with any nucleon transfer. Measurements of absolute cross sections on a number of targets between $^{16}$O and $^{208}$Pb were analyzed in a consistent way, with the cross sections reduced to spectroscopic factors through the distorted-wave Born approximation with global optical potentials. Across the 124 cases analyzed here, induced by various proton- and neutron-transfer reactions and with angular momentum transfer $\ell = 0$–7, the results are consistent with a quenching factor of 0.55. This is an apparently uniform quenching of single-particle motion in the nuclear medium. The effect is seen not only in $(d,p)$ reactions but also in reactions with $A = 3$ and 4 projectiles, when realistic wave functions are used for the projectiles.

PACS numbers: 21.10.Jx, 25.30.Dh, 25.40.Hs, 25.55.Hp

The mean-field description of nuclei, where valence nucleons move in single-particle orbits in the field generated by the remaining nucleons, has been tremendously successful. It is the basis of the shell model of nuclear structure, which quantitatively describes many aspects of nuclear properties. The mean field is, however, an approximation. At close distances, short-range correlations between nucleons must interfere with such a description. The study of processes where single nucleons are added or removed from a nucleus is a way of testing the limits of the approximation.

The data from transfer reactions form much of the experimental foundation of our understanding of nuclear structure. Cross sections from nucleon-transfer reactions are a measure of the overlaps between the target nucleus and the states formed when a nucleon is added or removed from the target. Thus these reactions have been used to test models of nuclear structure by comparing spectroscopic overlaps between initial and final nuclear states. The spectroscopic overlaps are represented by spectroscopic factors, effectively reduced cross sections. They are the experimentally measured cross section divided by the calculated one for a single-particle state with the same energy and quantum numbers. The summed reduced cross sections (or spectroscopic factors) with a given set of quantum numbers $\ell, j$ are a measure of the occupancy of the corresponding orbit [1].

In recent work [2, 3] the quantitative consistency of nucleon transfer – in particular, the reduction of experimental cross sections using the distorted-wave Born approximation (DWBA) – was investigated using cross sections from nucleon-transfer reactions on the stable Ni isotopes. According to the Macfarlane-French sum rules [1], the summed spectroscopic strengths, including both adding and removing on a given target, must be equal to the degeneracy of the orbit in question. This provides a method for determining the factor by which the observed cross sections, corrected for the reaction mechanism, differ from expectations.

It was found [2–4] that this procedure gave consistent and very similar normalizations for groups of nearby target nuclei. In the present Letter we focus our attention on the *value* of this normalization and extend our analysis to include reaction data on targets $16 \leq A \leq 208$. A variety of proton- and neutron-transfer reactions is considered, again with the criterion that the Macfarlane-French sum rules are satisfied. For reactions such as $(^3\text{He},\alpha)$ and $(\alpha,^3\text{He})$, we use the recent form factors based on Green's function Monte Carlo (GFMC) calculations [5] where previously more empirical approaches have been used.

In the late 1980s and early 1990s, careful studies were carried out at the NIKHEF facility of the $(e,e'p)$ proton-knockout reaction [6] on a range of nuclei from $A = 7$ to 208. Reactions with hadronic probes are localized to the nuclear surface, while the $(e,e'p)$ reaction probes the interior of the nucleus, mapping out the shape of the valence nucleon wave functions. The extraction of spectroscopic factors from this process is considerably less model dependent than for hadron-induced reactions. Concentrating mostly on closed-shell nuclei, Ref. [6] found that the spectroscopic factors of low lying states derived from $(e,e'p)$ for proton removal were consistently too low by a factor of almost two. A reanalysis of proton-removing $(d,^3\text{He})$ reactions on the same targets as those used in the $(e,e'p)$ studies, using consistent parameters in the analysis, showed a similar level of quenching [7]. Estimates of the effect of short-range correlations on these absolute overlaps by Pandharipande and co-workers concluded that "at any time only 2/3 of the nucleons in the nucleus act as independent particles moving in the nuclear mean field" [8].

In the past, the reduction of transfer cross sections to spectroscopic factors was done using DWBA to account for the energy- and target-dependence of the overlaps between the incident and outgoing channels and the final bound states. It was often done with different assumptions and approximations, and different distorting potentials and bound-state parameters. The resulting spectroscopic factors could only be considered in a relative sense, and while various normalization procedures were adopted in an attempt to determine the absolute scale, this was often not done consistently. Indeed, the need to measure accurate absolute cross sections, as opposed to relative ones, was often not a primary objective of many measurements. With faster computers, DWBA calculations were modified to include finite-range effects. There have been several new global surveys that parameterize the dependence of optical-model potentials on energy, mass, and proton-neutron ratio (Refs. [11, 25, 26] are examples).

In several recent experiments [2, 4, 12] we have made use of the normalization procedure described above, consistent within each experiment, but we were not consistent in the choice of the precise recipes for parameters between different analyses. In the present work we have re-analyzed these data in an overall consistent fashion, and included some of our other measurements. A few results from the literature, for which the cross sections seemed to have been measured carefully and where it was possible to retrieve the numerical values from the publications, have also been included.

The data considered here fall into two distinct categories, which were treated slightly differently. The first type involved data where both adding and removing reactions on the same target nuclei were available. For these data, the normalization was extracted by summing all of the adding and removing strength for a given $\ell, j$, and requiring that the sum of these adds up to the total degeneracy of the orbit. The associated quenching factor $F_q$ is given by

$$F_q \equiv \frac{1}{(2j+1)} \left[ \Sigma \left( \frac{\sigma_{\text{exp}}}{\sigma_{\text{DW}}} \right)_j^{\text{add}} + \Sigma \left( \frac{\sigma_{\text{exp}}}{\sigma_{\text{DW}}} \right)_j^{\text{rem}} \right], \quad (1)$$

where $\sigma_{\text{exp}}$ and $\sigma_{\text{DW}}$ are the experimental and DWBA cross sections, respectively. The superscripts 'add' and 'rem' denote adding and removing cross sections.

The second type were those where only adding or only removing data were available for a given nucleus. The above method was modified, with the assumption of a closed shell as was done for the $(e, e'p)$ data [7], to require that the total strength add up to the number of vacancies in the closed shell, or the number of particles outside it,

such that

$$F_q \equiv \frac{1}{(2j+1)} \left[ \Sigma \left( \frac{\sigma_{\text{exp}}}{\sigma_{\text{DW}}} \right)_j \right]. \quad (2)$$

The question of missing fragmented strength was addressed in Refs. [3] and [4], where it was found that the observed strengths of transitions, in between shells, were distributed in an approximately Lorentzian shape in excitation energy. The fraction estimated to fall outside the window of observation was less than a few percent of the total. It was also found in Ref. [3] that the spectroscopic factors for weak transitions begin to be unreliable for states that are lower than $\sim 0.1\%$ of the single-particle strength (for a well-matched $\ell$ value), probably because of competition from higher-order processes.

The majority of the data we have re-analyzed here is from experiments at the now-closed Wright Nuclear Structure Laboratory at Yale University between 2003 and 2011. Most of these data are published, though as mentioned, the analyses were not carried out with exactly the same parameters. All the measurements yielded absolute cross sections at the maxima of the angular distributions. The reactions were carried out at energies a few MeV/u above the Coulomb barrier. For this analysis, as we had done previously, momentum-matching considerations were taken into consideration: for all the transitions considered, $|k_{\text{out}} - k_{\text{in}}|R \approxeq \ell$ to within 1-2 $\hbar$, where the $k$'s are the incident and outgoing momenta (corrected for the Coulomb barrier), $R$ is an estimate of the average interaction radius, and $\ell$ is the orbital angular-momentum transfer.

The data analyzed here include the following adding and removing reactions performed on the same targets: $(d,p)$, $(p,d)$, $(\alpha,^3\text{He})$, and $(^3\text{He},\alpha)$ on $^{58,60,62,64}$Ni [2, 3], $^{74,76}$Ge, $^{76,78}$Se [4], and $^{130}$Te [12]. Those data for which just adding or removing reactions were probed include: $^{16}\text{O}(d,p)$ [13], $^{40}\text{Ca}(d,p)$ [14], $^{48}\text{Ca}(d,p)$ [15], the $(d,p)$ and $(\alpha,^3\text{He})$ reactions on $^{88}$Sr, $^{90}$Zr, and $^{92}$Mo [16], the $(d,p)$, $(p,d)$, $(^3\text{He},\alpha)$, and $(^3\text{He},d)$ reactions on $^{98,100}$Mo and $^{100,102}$Ru [17], the $^{112-124}\text{Sn}(^3\text{He},d)$ and $(\alpha,t)$ reactions [18], the $(p,d)$, $(^3\text{He},\alpha)$, and $(\alpha,^3\text{He})$ reactions on the stable even $N = 82$ isotones [19, 20], and the $(d,p)$ [21] and $(\alpha,^3\text{He})$ [22] reactions on $^{208}$Pb. In the case where there were no published numerical cross section values, angular-distribution plots were digitized and data extracted, with an increase in the uncertainties from this process estimated as $< 5\%$.

To allow for distortions and kinematic matching in relating cross sections to spectroscopic factors it is convenient to use the DWBA. The target single-particle wave functions of the transferred nucleon were generated using a Woods-Saxon potential with fixed geometrical parameters. This analysis was carried out using the finite-range DWBA code PTOLEMY [23] with the following considerations.

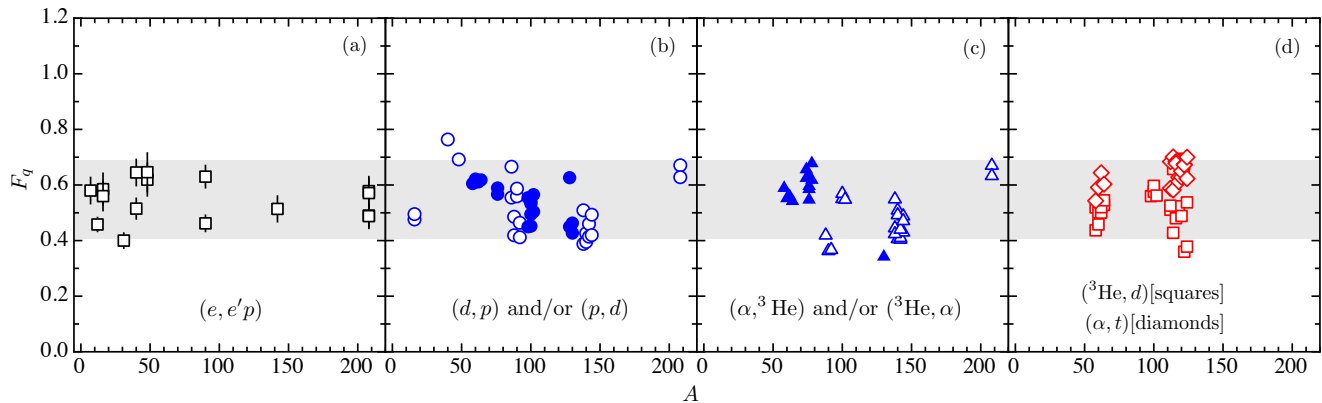

FIG. 1. The quenching factor $F_q$ versus target mass $A$. The $(e,e'p)$ data in Panel (a) are from Refs. [7, 9]. The grey band represents the mean $\pm 2\sigma$ of the $(e,e'p)$ data to guide the eye. The data in Panels (b), (c), (d) are from this analysis and are tabulated in the Supplemental Material [10]. Solid symbols are from adding and removing reactions while the empty ones are from just adding or just removing.

*Projectile wave function.*—For the deuteron wave function we used the Argonne $v_{18}$ potential [24] in the $(d,p)$ and $(p,d)$ reactions. For the more complex $A = 3$ and 4 projectiles, we used the GFMC-derived parameterizations from Brida et al. [5].

*Target wave function.*—The potential depth was varied to match the binding energy of the transferred nucleon for the state in question. The radial parameters were chosen to be consistent with the values obtained in the $(e,e'p)$ work of Ref. [7] with a radius parameter $r_0 = 1.28$ fm and diffuseness $a = 0.65$ fm, representing the average values. The spin-orbit potential depth was $V_{so} = 6$ MeV, with $r_{so0} = 1.1$ fm, and $a_{so} = 0.65$ fm.

*Optical-model potentials.*—For protons we used the global potentials of Koning and Delaroche [25]. Similarly for deuterons, we used the global potentials of Ref. [11], and for $^3$He, the recent study of Ref. [26]. The latter was also used for tritons, though it is less clear how appropriate it is. For $\alpha$ particles, we used the fixed potential of Ref. [27] that was derived from the $A = 90$ mass region. Other reasonable choices for potentials give similar results [2, 4].

The values of $\sigma_{\rm exp}/\sigma_{\rm DW}$ were used with Eq. (1) or (2) to obtain quenching factors $F_q$ that are summarized in Table I, categorized by reaction. A complete table of the data is in the Supplemental Material [10]. The quenching factors obtained in this analysis are also plotted in Fig. 1, along with those from $(e,e'p)$, as a function of mass number. The value appears to be independent of target mass and reaction, with a mean value of 0.55 and an rms variation of 0.10. It is also comparable to that seen in the $(e,e'p)$ data. Figure 2 shows the data emphasizing that the quenching factor is independent of $\ell$ value, at least between 0 and 7.

The uncertainties in the $F_q$ values are difficult to estimate. As noted previously (e.g. Ref. [2]), systematic effects dominate the uncertainties including errors in absolute cross sections, missed (or mis-assigned) states, the robustness of assumed shell closures, the effects of multistep mechanisms, and the choice of parameters in the DWBA analysis, and indeed in the assumptions inherent in DWBA. For a global average value for $F_q$ of 0.55 we find the rms variations amongst all the individual determinations to be 18%.

The only data that our group had obtained in the past decade that does not fit this pattern is a measurement with the $(d,^3{\rm He})$ reaction [28], taken at much higher energies than the rest of the results included here, $\sim$35 MeV/u above the Coulomb barrier instead of the $\sim$2-5 MeV/u for the rest. The value of $F_q$ obtained for the high energy data set, using the global optical-model potentials adopted in this analysis, was found to be internally consistent but $F_q \approx 1$ instead of 0.55. We found that at the higher energies, the sensitivity to the choice of optical-model distortions amongst various global parameterizations is much larger ($\sim$60%) than at the lower energies. For the rest of the data represented here, the cor-responding sensitivity for all reactions was $< 10\%$, apart from the $(^3\text{He},d)$ reaction which is $< 20\%$. The higher energy data are therefore not included in the present analysis. The sensitivity to parameters perhaps points to problems with the parameterizations in the global potentials for energies far above the barrier.

TABLE I. Mean quenching factor by reaction type.

| Reaction, $\ell$ transfer | Number of Determinations | $F_q$ | rms spread |
|---|---|---|---|
| $(e,e'p)$, all $\ell$ | 16 | 0.55 | 0.07 |
| $(d,p)$, $(p,d)$, $\ell = 0$-2 | 40 | 0.53 | 0.09 |
| $(d,p)$, $(p,d)$, $\ell = 0$-3 | 46 | 0.53 | 0.10 |
| $(\alpha,^3{\rm He})$, $(^3{\rm He},\alpha)$, $\ell = 4$-7 | 26 | 0.50 | 0.09 |
| $(\alpha,^3{\rm He})$, $(^3{\rm He},\alpha)$, $\ell = 3$-7 | 34 | 0.52 | 0.09 |
| $(^3{\rm He},d)$, $\ell = 0$-2 | 18 | 0.54 | 0.10 |
| $(^3{\rm He},d)$, $\ell = 0$-4 | 26 | 0.54 | 0.09 |
| $(\alpha,t)$, $\ell = 4$-5 | 14 | 0.64 | 0.04 |
| $(\alpha,t)$, $\ell = 3$-5 | 18 | 0.64 | 0.04 |
| All transfer data[a] | 124 | 0.55 | 0.10 |

[a] Rows 3, 5, 7, and 9.

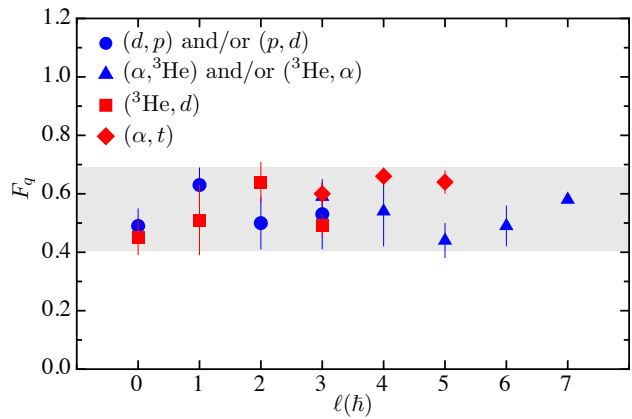

FIG. 2. Average of the quenching factor for different $\ell$ transfer. The error bars shown represent the rms spread in values. The grey band is the same as in Fig. 1.

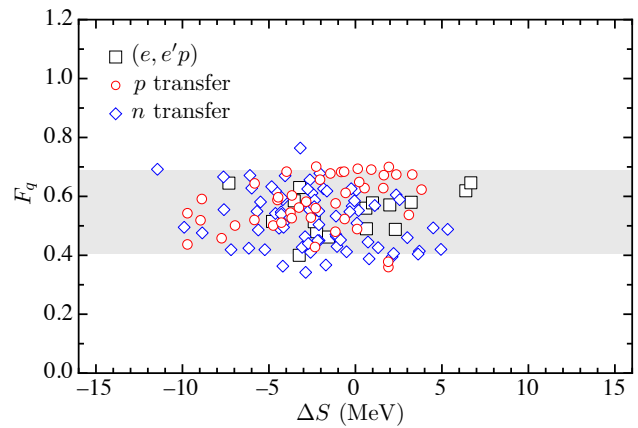

FIG. 3. $F_q$ versus $\Delta S$. The $(e,e'p)$ data are from Refs. [7, 9]. The grey band represents the $(e,e'p)$ data as in Fig. 1.

Gade *et al.* [29] plotted a 'reduction factor', which is the spectroscopic factor derived from measured cross sections divided by the expected shell-model value for a given state, versus an asymmetry parameter $\Delta S$ defined as $S_n - S_p$ (or $S_p - S_n$) for neutron knockout (or proton knockout). $\Delta S$ is therefore an approximate measure of the difference in the proton and neutron Fermi surfaces. Results from nucleon-knockout reactions have shown a trend, where this quantity approaches unity for large negative values of $\Delta S$, and becomes much smaller, around 0.2, for large positive values. However, Lee *et al.* [30] saw no such trend in $(p,d)$ transfer reactions on various Ar isotopes, though it has been suggested that the results may not be definitive [31]. In the recent work of Ref. [32], no such behavior in the reduction factor was found in proton- and neutron-removing from $^{14}$O, probing extreme positive and negative values of $\Delta S$. We display our results plotted against the more limited range in $\Delta S$ that is accessible with stable targets (about half what can be covered with radioactive beams) in Fig. 3, where no obvious trend is seen.

Other reaction models can be used to reduce experimental cross sections to spectroscopic overlaps, and one may perhaps expect that, if applied consistently, they are likely to yield similar results. For example, we used the finite-range adiabatic wave approximation formalism of Johnson and Tandy [33] with the code TWOFNR [34] for $\ell = 1$ $(p,d)$ and $(d,p)$ on the Ni isotopes. The values of $F_q$ differ by less than 10%. We used DWBA as the most convenient method to remove the dependence of the reaction cross sections on energy, nucleus, angular momentum, and reaction type.

The quenching of the single-particle mode appears to be a quantitatively uniform property of the nuclear many-body system from light to heavy nuclei. Correcting for this quenching makes the measured spectroscopic factors directly comparable to spectroscopic factors from shell-model calculations of nuclear structure that start with nucleons in independent-particle orbits. For models where many-body effects are taken into account, such as *ab-initio* calculations of nuclear structure, the correlations are already included, and spectroscopic overlaps may be directly compared to calculations (e.g. Ref. [35]).

In summary, we find that spectroscopic factors from single-nucleon transfer reactions derived from a self-consistent analysis are quenched with respect to the values expected from mean-field theory by a constant factor of 0.55, with an rms spread of 0.10, independent of whether the reaction is nucleon adding or removing, whether a neutron or proton is transferred, the mass of the nucleus, the reaction type, or angular-momentum transfer, at least in the range of stable nuclei.

The authors would like to thank S. C. Pieper and L. Lapikás for helpful discussions, as well as our experimental collaborators. This work was supported by the US Department of Energy, Office of Nuclear Physics, under Contract No. DE-AC02-06CH11357, and the UK Science and Technology Facilities Council.

* Electronic address: kay@phy.anl.gov
[1] M. H. Macfarlane and J. B. French, Rev. Mod. Phys. **32**, 567 (1960).
[2] J. P. Schiffer *et al.*, Phys. Rev. Lett. **108**, 022501 (2012).
[3] J. P. Schiffer *et al.*, Phys. Rev. C **87**, 034306 (2013).
[4] J. P. Schiffer *et al.*, Phys. Rev. Lett. **100**, 112501 (2008).
[5] I. Brida, Steven C. Pieper, and R. B. Wiringa, Phys. Rev. C **84**, 024319 (2011).

[6] L. Lapikás, Nucl. Phys. A **553**, 297c (1993).
[7] G. J. Kramer, H. P. Blok, and L. Lapikás, Nucl. Phys. A **679**, 267 (2001).
[8] Vijay R. Pandharipande, Ingo Sick, and Peter K. A. de-Witt Huberts, Rev. Mod. Phys. **69**, 981 (1997).
[9] L. Lapikás and H. P. Blok, private communication.
[10] The Supplemental Material will appear with the publication.
[11] Haixia An and Chonghai Cai, Phys. Rev. C **73**, 054605 (2006).
[12] B. P. Kay *et al.*, Phys. Rev. C **87**, 011302(R) (2013).
[13] J. L. Alty *et al.*, Nucl. Phys. A **97**, 541 (1967).
[14] L. L. Lee, Jr. *et al.*, Phys. Rev. **136**, B971 (1964) and erratum L. L. Lee, Jr. *et al.*, *ibid.* **138**, AB6 (1965).
[15] W. D. Metz, W. D. Callender, and C. K. Bockelman, Phys. Rev. C **12**, 827 (1975).
[16] D. K. Sharp *et al.*, Phys. Rev. C **87**, 014312 (2013).
[17] S. A. McAllister *et al.*, to be published.
[18] A. J. Mitchell, Ph.D. thesis, University of Manchester, 2012.
[19] A. M. Howard, Ph.D. thesis, University of Manchester, 2011.
[20] B. P. Kay *et al.*, Phys. Lett. B **658**, 216 (2008).
[21] A. F. Jeans *et al.*, Nucl. Phys. A **128**, 224 (1969).
[22] R. Perry *et al.*, Phys. Rev. C **24**, 1471 (1981).
[23] M. H. Macfarlane and S. C. Pieper, ANL-76-11 Rev. 1, ANL Report (1978).
[24] R. B. Wiringa, V. G. J. Stoks, and R. Schiavilla, Phys. Rev. C **51**, 38 (1995).
[25] A. J. Koning and J. P. Delaroche, Nucl. Phys. A **713**, 231 (2003).
[26] D. Y. Pang *et al.*, Phys. Rev. C **79**, 024615 (2009).
[27] G. Bassani and J. Picard, Nucl. Phys. A **131**, 653 (1969).
[28] B. P. Kay *et al.*, Phys. Rev. C **79**, 021301(R) (2009).
[29] A. Gade *et al.*, Phys. Rev. C **77**, 044306 (2008) and references therein.
[30] Jenny Lee *et al.*, Phys. Rev. Lett. **104**, 112701 (2010) and Jenny Lee *et al.*, Phys. Rev. C **83**, 014606 (2011).
[31] F. M. Nunes, A. Deltuva, and June Hong Phys. Rev. C **83**, 034610 (2011).
[32] F. Flavigny *et al.*, Phys. Rev. Lett. **110**, 122503 (2013).
[33] R. C. Johnson and P. C. Tandy, Nucl. Phys. A **235**, 56 (1974).
[34] J. A. Tostevin, University of Surrey version of the code TWOFNR (of M. Toyama, M. Igarashi, and N. Kishida) and code FRONT (private communication).
[35] L. Lapikás, J. Wesseling, and R. B. Wiringa, Phys. Rev. Lett. **82**, 4404 (1999).